\documentclass[prl,floatfix,showpacs,groupedaddress,twocolumn]{revtex4}
\usepackage{t1enc}
\usepackage[latin2]{inputenc}
\usepackage[american]{babel}
\usepackage{graphics,graphicx}
\usepackage{amsmath}
\usepackage{amssymb}
\usepackage{epsf}
\def\textbf#1{#1}
\begin{document}
\title{Resonance \textbf{enhancement} effects in Raman-enhancing \textbf{pyramid-like V-shape} groove microstructures}

\author{Matyas Mechler}

\affiliation{Department of Experimental Physics, University of P\'ecs, P\'ecs, Hungary}

\author{Sergei V.\ Kukhlevsky}

\affiliation{Department of Experimental Physics, University of P\'ecs, P\'ecs, Hungary}

\author{Adam Mechler}

\affiliation{School of Chemistry, Monash University Clayton, VIC 3800, Australia}

\author{Don McNaughton}

\affiliation{School of Chemistry, Monash University Clayton, VIC 3800, Australia}

\begin{abstract}
Microscopic pyramidal pits in a reflective surface, a geometry similar to a
retroreflector, are frequently used to enhance signal strength.
The enhancement effect is generally attributed to surface plasmons, however, \textbf{the
sub-wavelength to near-wavelength dimensions of the pyramidal 3D geometry}
suggest contributions from diffraction and near-field effects.
\textbf{Our} theoretical analysis of the light intensity distribution \textbf{in the similar (but simpler) 2D geometry} assuming a perfect conductor screen, that is, in the absence of any plasmon effects, shows that interference patterns forming within the cavity cause a significant \textbf{resonant} increase in local intensity. \textbf{Such effect can be important for many applications, especially for the widely used Raman spectroscopy.} 
Resonant enhancement \textbf{without plasmons} of the emitted Raman signal due to enhanced local field amplitude is also
possible, which implies that the geometry practically implements a Raman laser.
Comparison of diffraction patterns obtained with near-field and far-field approaches
reveals that the near-field component is responsible for the observed dramatic
intensity enhancement, and thus the Raman enhancement as well.

\end{abstract}

 \maketitle

\section{Introduction}

Micro- and nano-structured surfaces are frequently used to implement Surface Enhanced
Raman Scattering (SERS), where interaction of the analyte with the substrate,
mostly coupling between surface plasmons of the substrate and Raman modes of
molecules \cite {garcia96, perchec08},  lead to the multiplication of the Raman
signal intensity.  Controlling the SERS enhancement, however, has proven problematic
due to the random plasmon distribution.

Nanovoids such as spheric cavities or grooves are known to host well-defined plasmon
modes that interact strongly with incident laser light \cite{coyle01, perchec08}. It
was thus proposed that nanovoids might be used to achieve tunable Raman enhancement 
\cite{Netti07}. One of the geometries that performs particularly well consists of an
array of 1 $\mu$m deep pyramidal pits of $\sim$90\textdegree opening angle
\cite{klarite}. Remarkably, this mesoscale geometry has reflecting surfaces of a
distance that scales between a few wavelengths and sub-wavelength and thus resembles
an (unstable) optical resonator. This raises the possibility that in this particular
case the enhanced signal is generated by energy coupling into the Raman active medium
from a high energy standing wave, practically implementing a Raman laser.

While parallel mirror micro-resonators have been studied before \cite{klemens06}, the
tilted mirror geometry of the pyramidal pit requires a different formalism to account
for diffraction and near-field effects. Importantly, this geometry might be seen as a
special case of an aperture (with an infinitesimally small opening at the tip of the
pyramid). Light propagation through sub-wavelength apertures of various geometries is
intensively studied due to the special properties of the optical near field, in
particular, light localization and the accompanying intensity enhancement 
\cite{Betz1,Kuk2M,kuk6,Ebbe}. While experimental evidence prooves its
existence \cite{Ebbe}, the origin of the intensity enhancement is ambiguous as some
theories attribute it to the excitation of surface plasmon
polaritons \cite{Ebbe,Port,pendry2,krishnan}, others to intracavity (waveguide)
resonances \cite{baida,schouten,suzuki} or other phenomena \cite{garciapre,Gar1,Cao}.
While in the case of entirely sub-wavelength structures the comparable spatial
distribution of plasmon- and diffraction-originated intensities prohibits an easy
distinction, the dimensions of the pyramidal pit permit an attempt at the separation
of the plasmon and diffraction effects. 

When assuming that the aperture is in a perfect conductor screen, in which the
plasmon frequency is infinity, purely diffractive intensity distributions might be
calculated. 
The formulation we used is based on the near field treatment
by Neerhoff and Mur \cite{neerhoff,Betz1}.\textbf{We intend to find if resonant enhancement does exist in a pyramid-type geometry, therefore}
 we approximate the pyramidal pit with a V-slit, thus reducing the dimension of the problem to a plane, a simplification that,
at the moment, is necessary for the solution of the problem. Importantly, the V-slit
geometry is known to support standing waves as it can act as a waveguide; this
application was discussed in a number of studies. \cite{moreno,
novikov1,gramotnev1,pile1,pile2}

\section{Theory}
\begin{figure}[tb]
\centering
\includegraphics[width= 65 mm]{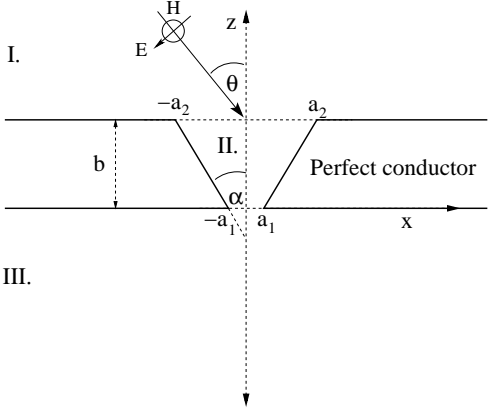}
\caption{Aperture model of the pyramidal pit.}\label{config}
\end{figure}

In this section we summarize briefly the formulation used. For further details on
a similar model see \cite{Betz1,Kuk2M}. The configuration is shown in
Fig.~\ref{config}. The slit is parallel to the $y$ axis in a screen of thickness $b$.
The entrance width of the slit is $a_2$, the exit width is $a_1$ (in our case,
$a_1\rightarrow 0$). The geometry is divided into three regions: I) $|x|<\infty$,
$|y|<\infty$, $b<z<\infty$, II) $|x|<a_1+\frac{a_2-a_1}{b}z$, $|y|<\infty$, $0<z<b$
and III) $|x|<\infty$, $|y|<\infty$, $-\infty<z<0$.

An incident plane wave propagating in the $x-z$ plane in region I arrives at the slit
at an angle $\theta$ with respect to the $z$ axis. The time harmonic magnetic field
is constant and at the same time polarized in the $y$ direction:
\begin{equation}
\label{mfield}
\mathbf{H}(x,y,z,t)=U(x,z)\exp(-i\omega t)\vec{e}_y.
\end{equation}
In this case the electric field can be found from Maxwell's equations, and the
diffraction problem is reduced to one involving the single scalar
field $U(x,z)$ in Eq.~\eqref{mfield} in only two dimensions.
In the three regions the field is represented by $U_j(x,z)$ ($j=1,2,3$). Due to time
harmonicity the field satisfies the Helmholtz equation. In region I the field is
composed of three components: $U_1(x,z)=U^i(x,z)+U^r(x,z)+U^d(x,z)$.
Here $U^i(x,z)$ is the incident field which is assumed to be a plane wave of unit
amplitude: $U^i(x,z)=\exp[ik_1(x\sin\theta-z\cos\theta)]$.  $U^r(x,z)$ is the field
that would be reflected without a screen: $U^r(x,z)=U^i(x,2b-z)$.  Finally,
$U^d(x,z)$ is the diffracted field produced by the presence of the slit.

For a perfectly conducting screen, the tangential $\mathbf{E}$ vanishes at the
surface $(\partial_n U=0)$. Further boundary conditions result from assuming that the
diffracted fields and their derivatives vanish at infinity. The final boundary
condition is the continuity of the field and its derivative at the upper and lower
slit boundaries. To find the field, the 2D Green's theorem is applied with one
function given by $U(x,z)$ and the other by a standard Green's function. We assumed
that the tangential $\mathbf{E}$ vanishes at the screen; therefore more is known
about $\partial_nU$ at the boundaries than is known about $U$ itself. The Green
functions for the V slit geometry can be found in \cite{morse}. The following
boundary conditions are imposed on $G$:
\small
\begin{subequations}
\begin{eqnarray}
\partial_z G_1(x,z)|_{z\to b^+}&=&0\quad \text{for }|x|<\infty,\\
\partial_z G_3(x,z)|_{z\to 0^-}&=&0\quad \text{for }|x|<\infty,\\
\partial_x G_2(x,z)|_{x\to \left(a_1+\frac{a_2-a_1}{b}z\right)^-}&=&0\quad
\text{for }0<z<b,\\
\partial_x G_2(x,z)|_{x\to \left(-a_1-\frac{a_2-a_1}{b}z\right)^+}&=&0\quad
\text{for }0<z<b.\quad
\end{eqnarray}
\end{subequations}
\normalsize
 $G_1$ and $G_3$ must also satisfy the Sommerfeld radiation condition. Using
these assumptions and conditions we may evaluate the three Green's functions. In
regions I and III we find
\begin{subequations}
\begin{eqnarray}
G_1(x,z;x',z')=\frac{i}{4}[H_0^{(1)}(k_1R)+H_0^{(1)}(k_1R')],\\
G_3(x,z;x',z')=\frac{i}{4}[H_0^{(1)}(k_3R)+H_0^{(1)}(k_3R'')],
\end{eqnarray}
\end{subequations}
where
\begin{subequations}
\begin{eqnarray}
R&=&\sqrt{(x-x')^2+(z-z')^2},\\
R'&=&\sqrt{(x-x')^2+(z+z'-2b)^2},\\
R''&=&\sqrt{(x-x')^2+(z+z')^2}.
\end{eqnarray}
\end{subequations}
In region II, the method of images can be used:
\begin{equation}
G_2=\pi i
\sum_{n=-\infty}^{\infty}\left\{H_0^{(1)}[k|\vec{r}-\vec{r_n'}|]+H_0^{(1)}[k|\vec{r}
-\vec{r_n''}|]\right\},
\end{equation}
where $\vec{r_n'}$ and $\vec{r_n''}$ define a series of images reflected by the
walls. The projection of an arbitrary point ($x',z'$) on the wall:
\begin{subequations}
\begin{eqnarray}
x_\times&=&\frac{z'+x'\tan\alpha+\frac{a_1}{\tan\alpha}}{\frac{1}{\tan\alpha}
+\tan\alpha},\\
z_\times&=&\frac{1}{\tan\alpha}\cdot(x_\times-a_1).
\end{eqnarray}
\end{subequations}
From this the coordinates of the mirrored point are $x_M=2x_\times-x'$ and $
z_M=2z_\times-z'$. A series of images of the original point might be generated with
the rotation of the points by $4n\alpha$ ($n$ being the number of rotation).

By using the definitions of the Green functions in the three regimes and the boundary
conditions, one can find the functions $U^d$, $U_2$, and $U_3$:
\small
\begin{subequations}
\label{eq:Uexps}
\begin{eqnarray}
U^d(x,z)&=&-\intop_{-a_2}^{a_2}\frac{\epsilon_1}{\epsilon_2}G_1(x,z;x',b)DU_b(x')dx'
\nonumber\\ &&\text{for } b<z<\infty,\\
U_3(x,z)&=&\intop_{-a_1}^{a_1}\frac{\epsilon_3}{\epsilon_2}G_3(x,z;x',
0)DU_0(x')dx'\nonumber\\ && \text{for } -\infty<z<0,\\
U_2(x,z)&=&-\intop_{-a_1}^{a_1}\big[G_2(x,z;x',0)DU_0(x')\nonumber\\
&&-U_0(x')\partial_{z'}G_2(x,z;x',z')|_{z'\to0^+}\big]dx'\nonumber\\
&&+\intop_{-a_2}^{a_2}\big[G_2(x,z;x',b)DU_b(x')\nonumber\\
&&-U_b(x')\partial_{z'}G_2(x,z;x',z')|_{z'\to b^-}\big]dx' \nonumber\\ 
&&|x|<a_1+\frac{a_2-a_1}{b}z \text{ and } -\infty<z<0,
\end{eqnarray}
\end{subequations}
\normalsize
where the boundary fields are the same as in \cite{Betz1}:
\begin{subequations}
\begin{eqnarray}
U_0(x)&=&U_2(x,z)|_{z\to0^+},\\
DU_0(x)&=&\partial_zU_2(x,z)|_{z\to0^+},\\
U_b(x)&=&U_2(x,z)|_{z\to b^-},\\
DU_b(x)&=&\partial_zU_2(x,z)|_{z\to b^-}.
\end{eqnarray}
\end{subequations}
By applying the continuity equations to these functions, a set of four integral
equations can be found. We solved the equations for $U^d$, $U_2$, and $U_3$
numerically.
\begin{figure}[tb]
\centering
\includegraphics[trim= 10 35 10 30 , width=58mm ]{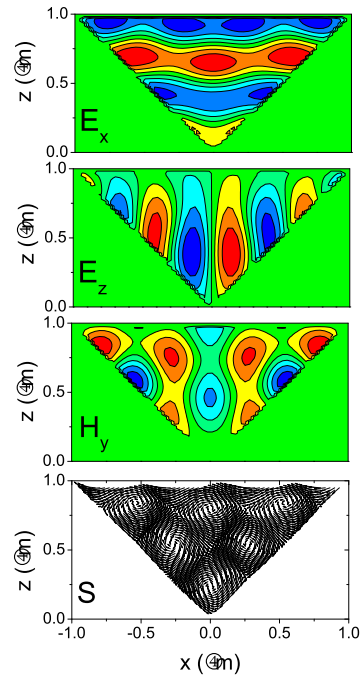}
\caption{$E_x$ and $E_z$ electric and $H_y$ magnetic field distributions, and the
electric field vector in the V-groove for $\lambda=534$~nm.}\label{szamolasok1}
\end{figure}

\section{Results and Discussion}

We analysed the light diffraction  in region II for several laser wavelength in the
range that is frequently used for Raman spectroscopy. A typical result is shown on
Figure \ref{szamolasok1} for $\lambda=534$~nm. Similar resonant intensity enhancement
was observed for all wavelengths. This enhancement is due to redistribution of
electric charges (i.e.\ not plasmons) on the screen surface leading to standing wave
formation with enhanced amplitude. The presence of the standing waves confirms that
the geometry acts as an optical resonator.

\begin{figure}
\centering
\includegraphics[
width= 0.5\textwidth]{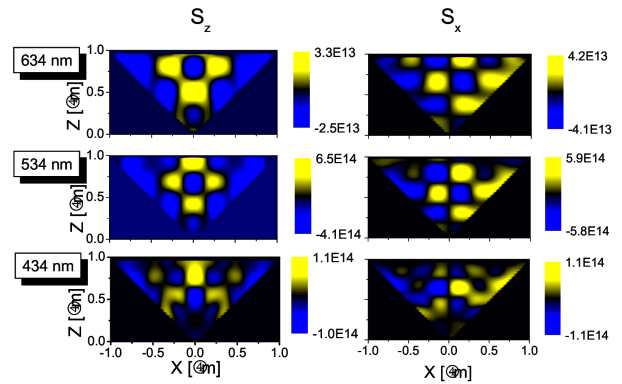}
\caption{\label{vectors}Electric field vectors for three wavelengths: 434~nm, 534~nm
and 634~nm.}
\end{figure}

Another interesting feature is the formation of vortices in the cavity (Figure
\ref{vectors}). These vortices appear approximately at $z\approx n\lambda/2$,
where $n=1,2,3,\dots$ It is important to note that at the surface the
electric field vectors are mostly parallel to the walls. Thus, when considering
the interaction with the Raman active material, the enhancement of different modes
should occur in comparison to plasmon-enhancement where the electric field is mostly
perpendicular to the surface. 
\begin{figure}
\centering
\includegraphics[width= 0.5 \textwidth ]{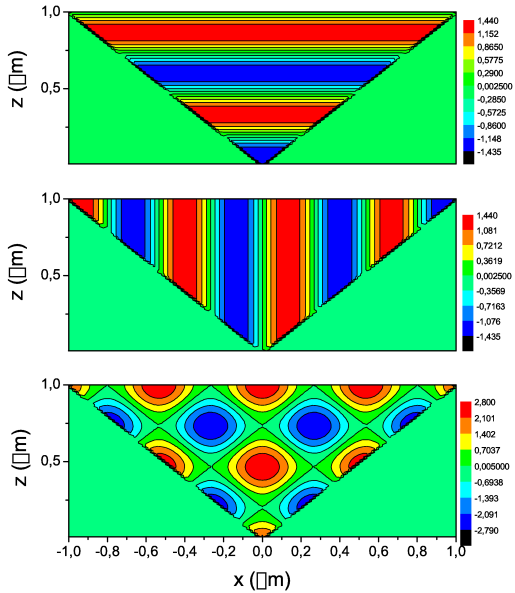}
\caption{\label{farfield}$E_x$ and $E_z$ electric and $H_y$ magnetic field distributions (far-field theory) $\lambda=534$~nm.}
\end{figure}

\begin{figure*}[tb]
\centering
\includegraphics[trim= 10 38 10 35 , width= 135 mm]{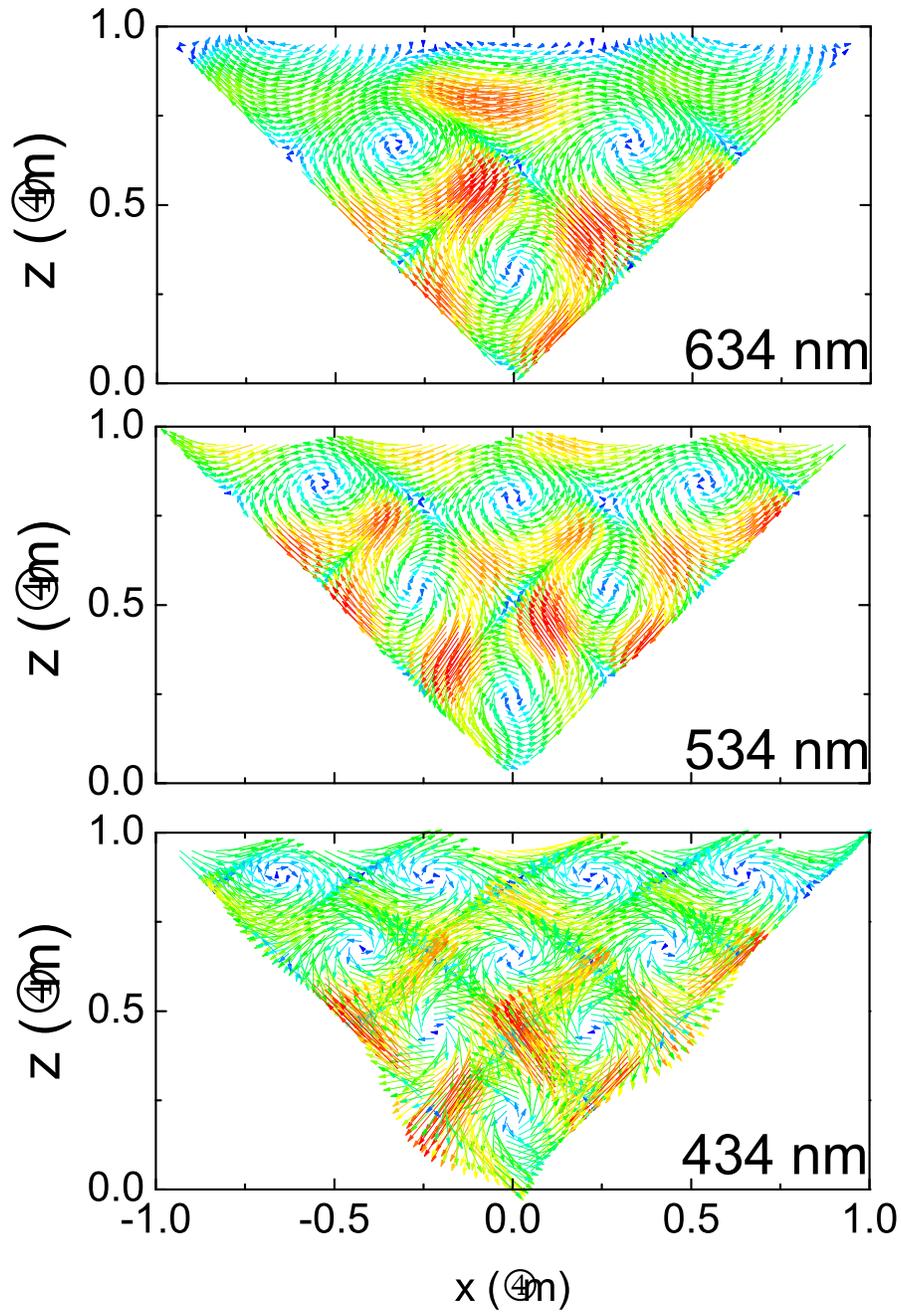}
\caption{Poynting vector for three wavelengths: 434~nm, 534~nm and 634~nm. Note the
``moving'' of the standing wave.}\label{poynting}
\end{figure*}

Far-field theory also predicts the standing waves and the vortices \textbf{(Fig.~\ref{farfield})},
however,  a few remarkable differences justify the use of near-field theory. First,
at the slit entrance ($z\to b^-$) a strong intensity suppression can be observed (see
Figs.~\ref{poynting} and \ref{vectors}) that cannot be seen in the case of a
macroscopic slit with the same geometry \textbf{(see Fig.~\ref{farfield})}. Notably, far-field diffraction theory used
to calculate intensity distributions in macroscopic objects does not contain the
diffractive component that causes the suppression. Second, the maximum electric
field occurs at the apex of the V-shape that confirms the observations of Suzuki
\textit{et al} \cite{suzuki}.

A comparison of the energy flux: $x$ and $z$ components of the Poynting vector are
depicted in Figure~\ref{poynting}. It is important to note that, while the standing
waves appear at distances that satisfy the $n\lambda/2$ criterium for the consecutive
wavelengths, the values of intensity maxima of the standing waves are strongly
wavelength dependent. This dispersion effect might influence the efficiency of
Raman-enhancement in these structures when using different laser wavelengths, and
could be used for the experimental verification of the predictions of the
calculations. Note that enhancement at different wavelengths is not the
same for electric field components and Poynting vector components, e.g.\ Poynting
vector shows notable enhancement for $\lambda=534$~nm in comparison with
$\lambda=434$~nm or $\lambda=634$~nm while for the same wavelength the maximal value
of the electric field vector is the smallest. 

In summary, theoretical analysis of light diffraction in a sub-wavelength to
near-wavelength V-groove \textbf{in a perfect conductor screen} was performed by using near-field approach. It was shown
that the geometry acts as an optical resonator where the intensity enhancement is
determined by the near-field effects\textbf{, i.e. without plasmons}.  The field vectors form vortices around the
intensity maxima, being mostly parallel to the walls of the groove, which determines
the Raman modes of surface bound molecules enhanced in this
configuration. The intensity enhancement is wavelength dependent, thus
different laser wavelengths might lead to different Raman enhancement when using this
geometry. Comparison of results obtained with near-field and far-field
approaches reveals that the near-field component is responsible for the observed
dramatic intensity enhancement, and thus the Raman enhancement as well.

\textbf{Acknowledgement}
AM acknowledges the Australian Research Council and the support of Monash University
through the Monash Fellowship scheme. This study was supported in part by the
Framework for European Cooperation in the field of Scientific and Technical Research
(COST, Contract No. MP0601) and the Hungarian Research and Development Program (KPI,
Contract No. GVOP 0066- 3.2.1.-2004-04-0166/3.0).

\bibliographystyle{apsrev}

\end{document}